\documentclass[review]{elsarticle}

\usepackage{lineno,hyperref}
\usepackage{float}
\usepackage{outlines}
\usepackage{hhline}
\modulolinenumbers[5]

\journal{Journal of \LaTeX\ Templates}

\begin{document}

\begin{frontmatter}

\title{Intra-Domain Task-Adaptive Transfer Learning to Determine Acute Ischemic Stroke Onset Time}

\author[affiliate1,affiliate2]{Haoyue Zhang\fnref{footnote1}}
\author[affiliate1,affiliate2]{Jennifer S Polson\fnref{footnote1}}
\author[affiliate3]{Kambiz Nael}
\author[affiliate3]{Noriko Salamon}
\author[affiliate3]{Bryan Yoo}
\author[affiliate4]{Suzie El-Saden}
\author[affiliate5]{Fabien Scalzo}
\author[affiliate1,affiliate3]{William Speier}
\author[affiliate1,affiliate2,affiliate3,affiliate6]{Corey W Arnold\corref{mycorrespondingauthor}}
\fntext[footnote1]{Haoyue Zhang and Jennifer S Polson contributed equally. }
\fntext[footnote2]{This work was supported by the United States National Institutes of Health (NIH) grants R01NS100806 and T32EB016640, and an NVIDIA Academic Hardware Grant.}
\cortext[mycorrespondingauthor]{Corresponding author}

\address[affiliate1]{Computational Diagnostics Lab, University of California, Los Angeles, CA 90024 USA}
\address[affiliate2]{Department of Bioengineering, University of California, Los Angeles, CA 90024 USA}
\address[affiliate3]{Department of Radiology, University of California, Los Angeles CA 90024 USA}
\address[affiliate4]{Department of Radiology, VA Phoenix Healthcare system, AZ 85012 USA}
\address[affiliate5]{Departments of Neurology and Computer Science, University of California, Los Angeles, CA 90024 USA}
\address[affiliate6]{Department of Pathology, University of California, Los Angeles CA 90024 USA}


\begin{abstract}
\label{sec:abstract}
Treatment of acute ischemic strokes (AIS) is largely contingent upon the time since stroke onset (TSS). However, TSS may not be readily available in up to 25\% of patients with unwitnessed AIS. Current clinical guidelines for patients with unknown TSS recommend the use of MRI to determine eligibility for thrombolysis, but radiology assessments have high inter-reader variability. In this work, we present deep learning models that leverage MRI diffusion series to classify TSS based on clinically validated thresholds. We propose an intra-domain task-adaptive transfer learning method, which involves training a model on an easier clinical task (stroke  detection) and then refining the model with different binary thresholds of TSS. We apply this approach to both 2D and 3D CNN architectures with our top model achieving an ROC-AUC value of 0.74, with a sensitivity of 0.70 and a specificity of 0.81 for classifying TSS $<$ 4.5 hours. Our pretrained models achieve better classification metrics than the models trained from scratch, and these metrics exceed those of previously published models applied to our dataset. Furthermore, our pipeline accommodates a more inclusive patient cohort than previous work, as we did not exclude imaging studies based on clinical, demographic, or image processing criteria. When applied to this broad spectrum of patients, our deep learning model achieves an overall accuracy of 75.78\% when classifying TSS $<$ 4.5 hours, carrying potential therapeutic implications for patients with unknown TSS.
\end{abstract}
\end{frontmatter}

\section{Introduction}
\label{sec:introduction}
Acute ischemic stroke (AIS) is a cerebrovascular disease accounting for 2.7 million deaths worldwide every year \cite{stroke_stats}. Treatment of AIS is heavily dependent on the time since stroke onset (TSS); current clinical guidelines recommend thrombolytic therapies for AIS patients presenting within 4.5 hours and endovascular thrombectomy for those presenting up to 24 hours after onset. AIS without a clear TSS is relatively common, accounting for up to 25\% of all AIS \cite{wakeup_ct, alteplase_45}. Some reasons for unclear TSS include unwitnessed strokes, wake-up strokes, or unreliable reporting by patients. For this patient population, the most recent AHA guidelines recommend using MRI sequences to assess patient eligibility for thrombolytics \cite{AHA_guidelines}. 

\begin{figure}[!t]
\centerline{\includegraphics[width=\columnwidth]{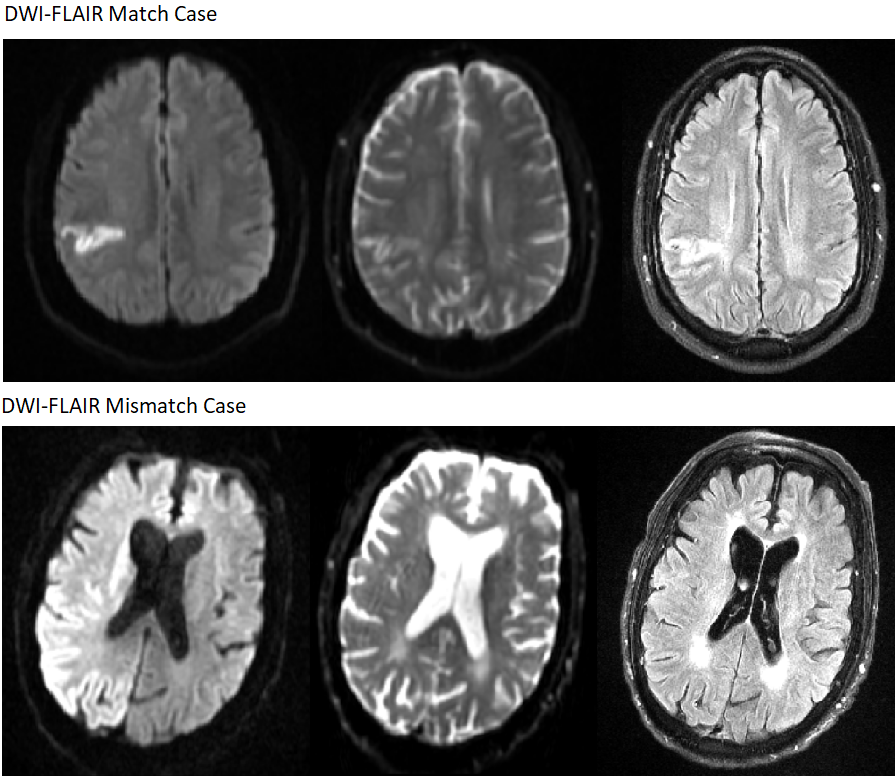}}
\caption{Sample cases of DWI-FLAIR Mismatch. Sequences from left to right: DWI b1000, DWI B0, FLAIR} \label{Figure 1}
\end{figure}

Following the WAKE UP trial \cite{MRIGuidedTF}, which used DWI-FLAIR mismatch to select patients for extending the time window for intravenous thrombolysis, the use of MRI (FLAIR-DWI mismatch) is now recommended (level IIa) to identify unwitnessed AIS patients who may benefit from thrombolytic treatment \cite{AHA_guidelines}. Specifically, diffusion-weighted imaging (DWI) displays increased signal in ischemic areas within minutes of stroke occurrence, while fluid-attenuated inversion recovery (FLAIR) imaging can show fluid accumulation after a few hours \cite{unknown_onset}, as shown in Fig. \ref{Figure 1}. A DWI-positive, FLAIR-negative mismatch can identify stroke lesions that could benefit from administration of thrombolytics. However, assessing this mismatch is subject to high variability compared across multiple readings and/or radiologists \cite{dwi-flair}. Thus, determining stroke onset using imaging alone could increase the number of patients eligible to receive thrombolytic treatments, possibly improving their outcomes. 

Several machine learning approaches have been used to determine stroke onset time in an automated fashion. These involve generating hand-crafted, radiomic, or deep learning-derived features from either clinical reports or images and then using these features as inputs to a variety of machine learning models \cite{tss_kch_1,tss_key,tss_korea}. This feature extraction has typically relied on defined regions of interest, determined by applying an image threshold or using a parameter map. Limiting features to this immediate region may fail to capture imaging characteristics in the surrounding region, which could be crucial to informing TSS given the interconnected nature of cerebral blood flow \cite{collaterals}. Moreover, previously published approaches have applied meticulous exclusion criteria, either by stroke location or imaging factors related to preprocessing; for these studies, as many as 40\% of patients were ineligible for assessment \cite{tss_korea}. 

Deep learning models have excelled in medical imaging for segmentation and classification tasks \cite{cnn_detection,segmentation,image_analysis, brain_tumor, isles}. Specifically, convolutional neural networks (CNNs) have produced state-of-the-art results even in small datasets common in medical imaging research \cite{small_dataset}. Convolutions, which aggregate pixel neighborhoods across layers, may occur in either two or three dimensions. While there has been a wide range of 2D CNNs applied to medical image tasks, 3D CNNs offer the added advantage of integrating information along the z-axis as well. The potential advantages of 3D convolutions come with a cost of increased model complexity, which generally requires a higher amount of data and computation power to train.

Due to the large number of parameters in a deep neural network, a high volume of data is typically required for training. For particularly complex classification tasks, transfer learning has been shown to achieve model convergence using less computation and boost performance in less time compared to training models from scratch \cite{transfer}. Transfer learning traditionally involves training a model on one dataset, then refining the model on another set of data for a different task. Cross-domain transfer learning involves training on data from a source domain, and using those learned weights in a model trained on data from a different target domain \cite{transfer_survey}, e.g., from the natural image domain to the medical image domain or from the CT image domain to the MR image domain. Many deep learning approaches applied to medical images have used established architectures pretrained on large natural image datasets such as ImageNet \cite{ImageNet} and refined the model to the domain-specific task. This is thought to improve model convergence, and use the low-level features learned on a high volume dataset for a smaller dataset, which is usually the case for medical image models given the high cost to acquire sufficient data. However, the differences in natural images and those in the medical domain limit the wide applicability of this method, likely due to over-parameterization of the original models \cite{transfer_review}. Efforts have been made to pretrain models on public medical datasets, but access to such medical datasets is still limited. Moreover, higher-level features of medical images vary significantly for different medical domains. To combat the limitations of cross-domain transfer learning and increase features reuse across models, intra-domain transfer learning has been implemented for both natural image and medical image tasks \cite{medical_transfer}. Commonly, a model is initialized in a self-supervised or unsupervised fashion. The advantage of this approach is that it does not require outside datasets or labels. However, even intra-domain pretraining may result in limited feature reuse beyond the first convolutional layer \cite{feature_reuse}. A task-adaptive approach, which uses the same data set for pretraining and then refines the model using two different label sets, has been demonstrated to increase feature reuse and enhance performance \cite{Elman1993,bengio}. However, this has not yet been applied in the medical image domain.

We propose an intra-domain task-adaptive transfer learning approach and implement it for TSS classification. The approach uses a multi-stage training schema, leveraging features learned by training on an easier task (stroke detection) to refine the model for a more difficult task (TSS classification). We developed both 2D and 3D CNN models to classify TSS, and we  demonstrated our proposed transfer learning approach enhanced classification performance for both architectures when compared to other pretraining schemas, with our 2D model achieving the best performance for classifying TSS $<$ 4.5 hours. We also showed that adding soft attention mechanisms during latter stages further improved the performance. To offer clinical insight, we compared our model performance to both previously published methods and radiologist assessment of DWI-FLAIR mismatch. Our deep learning models were able to achieve greater classification sensitivity while maintaining specificity achieved by expert neuroradiologists. By visualizing network gradients via Grad-CAM \cite{Gradcam}, we illustrated that our pretrained models were able to localize the stroke infarct more precisely than the models trained from scratch. To our knowledge, this is the first end-to-end, deep learning approach to classify TSS on a patient dataset with minimal exclusion criteria; moreover, our model exceeds the performance of previously reported state-of-art machine learning models.

\section{Material and Methods}

\subsection{Dataset and preprocessing}

A total of 422 patients treated for AIS at the UCLA Ronald Reagan Medical Center from 2011-2019 were included in this study. This work was performed under the approval of the UCLA Institutional Review Board (\#18-000329). A patient was included if they were diagnosed with AIS, had a known stroke onset time, and underwent MRI prior to any treatment, if given. Clinical parameters were gathered from imaging reports and the patient record, with demographic data summarized in Table \ref{tab:demographics}. The study cohort had a median age of 70 (55-80) years, a mean National Institutes of Health Stroke Scale (NIHSS) score of 8(4-15), and were 56\% female. The median onset to MRI was 222(105-715.25) minutes. For performance evaluation, we used 64\% for training (272), 16\% for validation (68), and 20\% (82) as a hold-out test set. In order to prevent information leakage across tasks, the same test set was used across a set of experiments. The training and testing sets had similar distributions of these clinical factors and TSS. For each patient in the test cohort, DWI-FLAIR mismatch was assessed independently by three senior neuroradiologists.

\begin{figure}[!t]
\centerline{\includegraphics[width=\columnwidth]{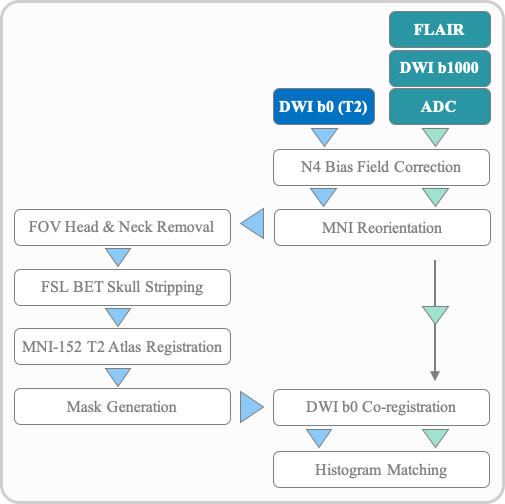}}
\caption{Preprocessing pipeline for patient series.} \label{preprocessing}
\end{figure}

\begin{table}[htbp]
  \centering

    \begin{tabular}{|l|l|l|}
    \hline
          & Training Set & Test Set \\
          & (n =340) & (n = 82) \\
          \hline
    Age (years) & 70 (55-80) & 68 (57-79) \\
    Female & 176 (52\%) & 46 (56\%) \\
    NIHSS & 8 (4 - 16) & 6.5 (2 - 18) \\
    Onset to MRI (min) & 210 (105-683) & 230 (107-661) \\
    \hline
    \end{tabular}
\caption{\label{tab:demographics}Patient cohort demographics. Numbers are n (\%) or median (interquartile ranges). MRI indicates magnetic resonance imaging; NIHSS, National Institutes of Health Stroke Scale.
}
\end{table}

\begin{figure}[!h]
\centerline{\includegraphics[scale=0.45]{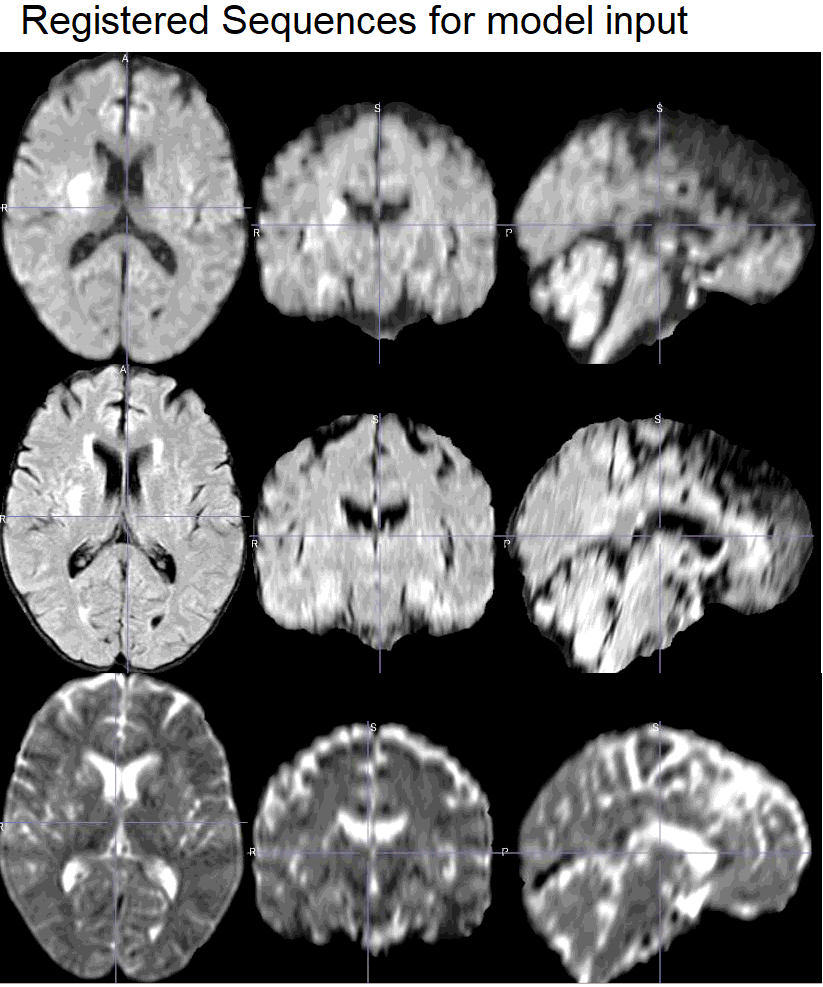}}
\caption{Sample case of Registered Output. Sequences from top to bottom: DWI(b1000), FLAIR, T2w(DWI B0).} \label{registeredsample}
\end{figure}

For each patient, the T2w(DWI B0), DWI(b1000) and FLAIR imaging sequences were retrieved from the institutional picture archiving and communication system (PACS). The sequences were then fed into our automated preprocessing pipeline. First, N4 bias field correction \cite{N4} was applied to all sequences. Then, each image series was reoriented to the T2w MNI-152 atlas \cite{mni152}. Next, the neck and skull were removed using FSL BET \cite{FSL}. The T2 sequence was registered using FSL FLIRT to a version of the T2w MNI-152 atlas that was down-sampled in order to match the z dimension of the stroke sequences. After a second run of FSL BET was performed to remove any artifacts, the remaining sequences were co-registered to the T2 volume. Finally, intensity was normalized, and histogram matching was performed using a reference study. This data preprocessing pipeline is summarized in Fig. \ref{preprocessing} and a sample output is in Fig. \ref{registeredsample}.

\subsection{Training Schema}

\begin{figure*}[h]
\centerline{\includegraphics[width=\textwidth]{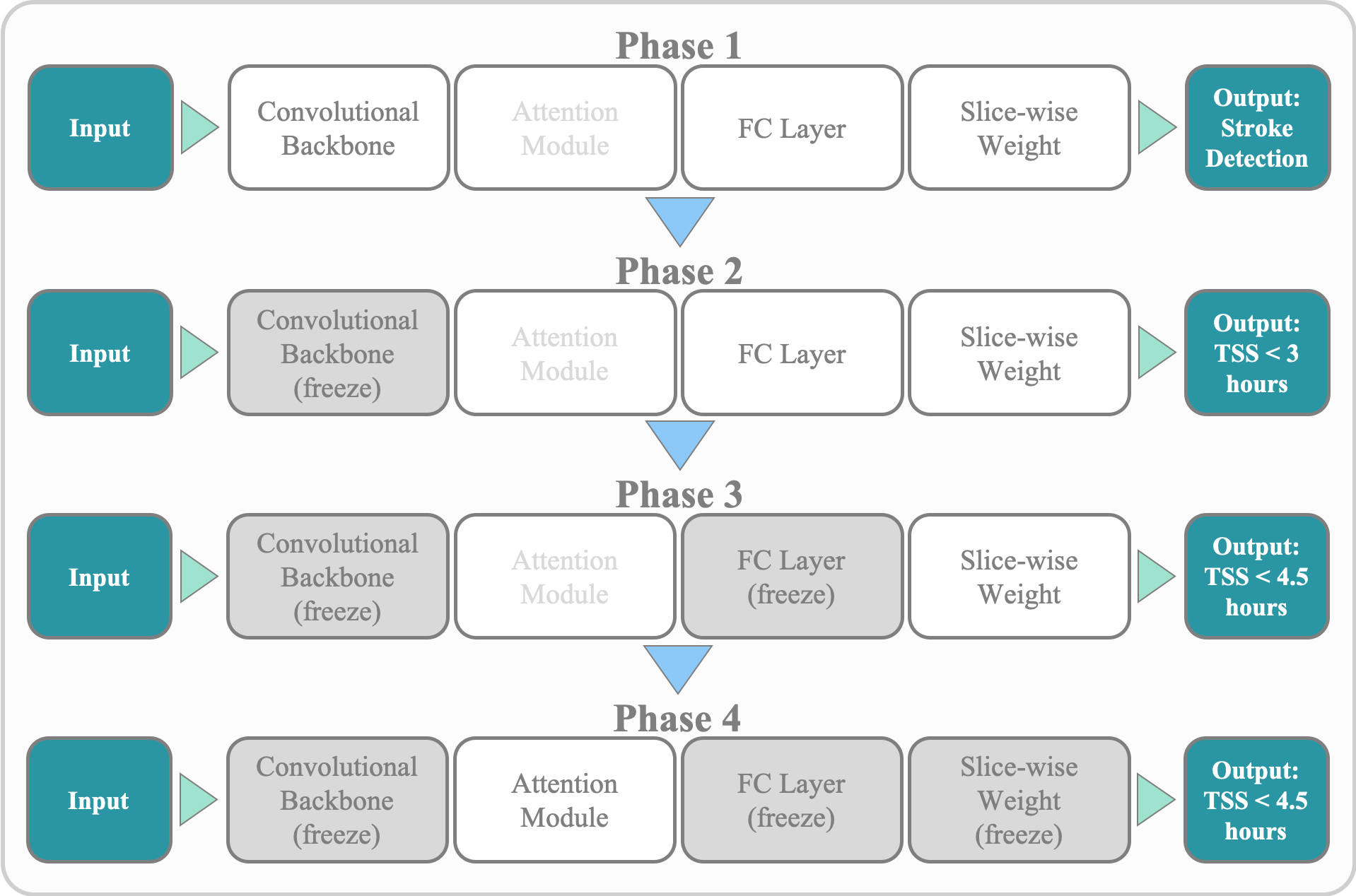}}
\caption{A summary of our training schema. Each phase utilized a unique classification label, as enumerated in the Outputs boxes for each phase. At the end of each training phase, the weights of certain components were frozen; these frozen weights were then initialized for the model at the start of the following phase.}
\label{schema}
\end{figure*}

To train the models, each brain volume was split into longitudinal halves. For each half, three imaging series, T2, DWI, and FLAIR, were concatenated and input as channels. The right halves were flipped on the vertical axis in order to spatially align with the left halves for inputs. Our models used a multi-phase training regimen. The first phase consisted of stroke detection, where brain halves were fed into the model separately and were labeled as positive (1) if they had a stroke lesion in the image. The 2D model was trained from scratch on this task. For our 3D model, initial weights were generated in a self-supervised fashion before the stroke side detection task for more rapid convergence. Once the model finished training, the first two convolutional layers/blocks were frozen. This pretrained network was then utilized in a second phase of training, whereupon only images with stroke lesions were used as input. In the second phase, we froze early convolutional weights to refine later layers and trained our model on TSS $<$ 3 hours, given the clinical correlation of DWI-FLAIR mismatch to this binarization. For the third phase, we used the pretrained weights of the TSS $<$ 3 hours model to train on the TSS $<$ 4.5 hours task. The last phase of our training schema (Fig. \ref{schema}) involved fine-tuning the soft attention modules to further enhance performance. We compared this multi-phase training regimen to training on TSS labels from scratch, pretraining on natural images, and pretraining on external datasets of brain MRIs \cite{2d_tumor_class_pretrained,3d_tumor_seg_pretrained}.

\subsection{Model Architectures}
We tested our intra-domain transfer learning schema on custom 2D and 3D architectures. The 2D CNN takes individual slices as input and feeds them through a convolutional backbone adapted from \cite{resnet} for feature extraction. To account for the large pixel input of an individual MRI slice, we also incorporated a soft attention gate into the architecture \cite{attention_2d}. This module uses the final and penultimate convolutional outputs to generate individual pixel weights which identify the most salient regions for the task. This attention module was refined during the TSS tasks later in training to avoid the possibility of convergence at a local minimum and precluding further optimization during model refinement \cite{attention_unet}. The attention module output and convolutional output were concatenated into a feature vector, which was then fed into a fully-connected layer to generate a single, slice-level output. To aggregate these slice-level predictions into an image-level prediction, we implemented a trainable weighting factor, ranging from 0 to 1, to assign a weight to each slice, and the slice-level outputs were summed in a weighted fashion, resulting in one probability label. The attention module and trainable weight factor ascribe pixel-level and slice-level importance that can be trained and optimized, which enables the model to localize to salient regions.

\begin{figure*}[!b]
\centerline{\includegraphics[width=\textwidth]{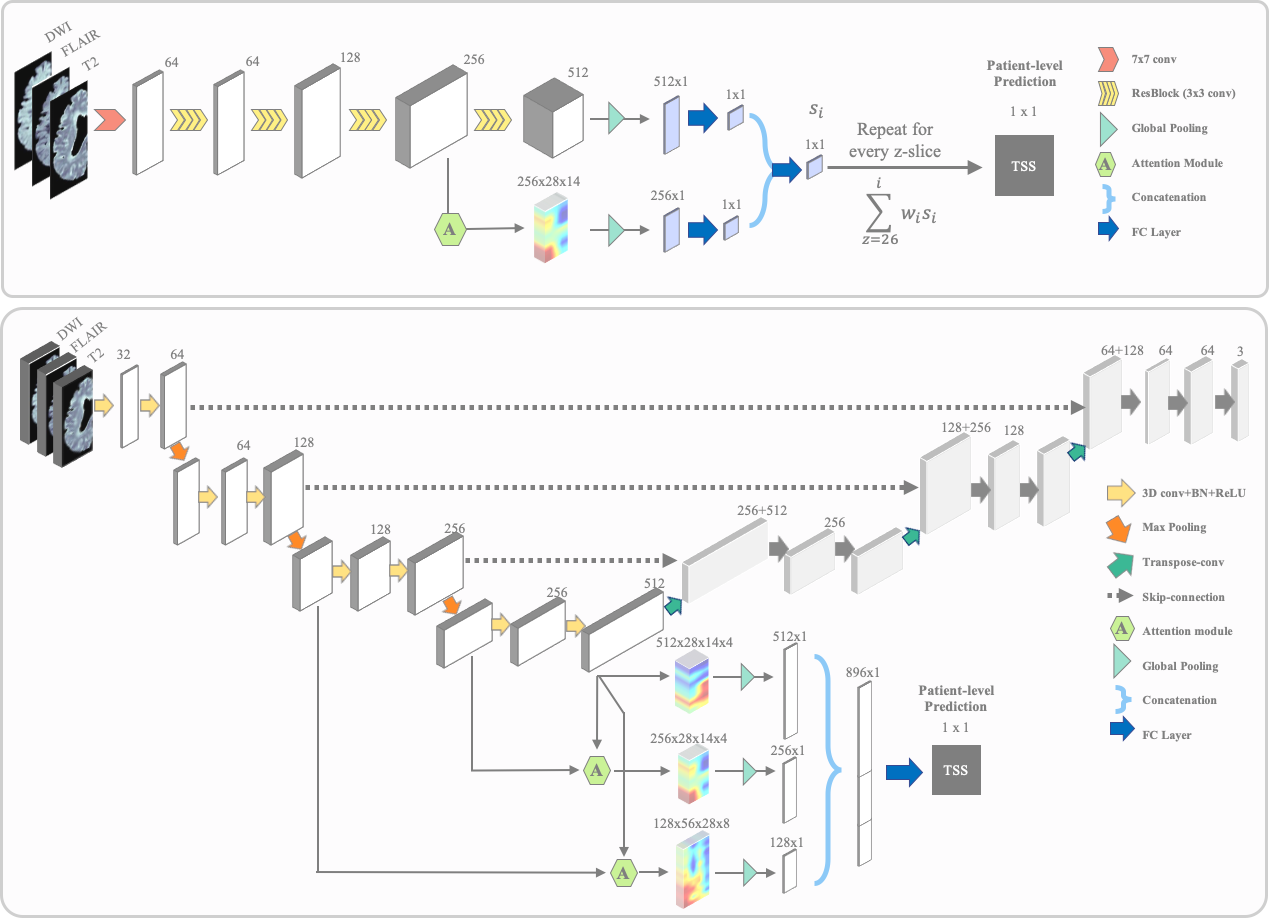}}
\caption{Architectures for 2D (top) and 3D (bottom) models. Our 2D Self-weighted Slice-wise Attention model took DWI b1000, B0, and FLAIR as a 3-channel input to a feature extraction backbone. Each slice of the brain was individually fed through 4 ResBlocks of ResNet-18 to generate a 512x7x4 feature map, then pooled to a 512x1 feature vector \cite{resnet}. A soft attention module at the 256-channel convolutional layer was added to generate a 256x28x14 attention feature map and then pooled to a 256x1 feature vector. The feature map and attention feature map were aggregated for each slice with a learnable weighting factor for final classification. Our 3D model first used the entire structure of a 3D U-Net to train an initial weight using Models Genesis. Then volumetric DWI, T2 and FLAIR were directly fed into the encoder part of the network. Two soft attention modules were added at 128 and 256-channel convolution layers. Feature maps from the original network and the two attention modules were pooled globally and concatenated for classification.} \label{arch}
\end{figure*}

Given the 3D anatomical information in our dataset, we also evaluated a 3D model architecture. Training a 3D CNN model from scratch does not necessarily yield better performance than 2D models due to the higher number of parameters and the potential for over-fitting. To address these challenges, we first adapted a self-supervised learning approach, known as Models Genesis \cite{model_genesis}, to train a 3D U-Net \cite{3d_unet} in order to generate initial weights for the stroke detection task; U-net \cite{2d_unet}, like ResNet, uses connections between layers for model training, and also has been widely used in medical image research. Our 3D approach also used soft attention modules at the 128- and 256-channel intermediate outputs in order to allow the network to capture relevant information in early stages of classification. Additionally, using Models Genesis, we first modified the original image and then trained a 3D U-Net to restore the original image, enabling the model to learn important high level features in the original image. We then used the encoder component of the 3D U-Net network, along with two soft attention modules, to train the model to classify stroke side and TSS. Fig. \ref{arch} illustrates the 2D Self-weighted Slice-wise Attention Model structure, and the 3D Attention Model structure. The Models Genesis and soft attention modules bolstered 3D model performance. 

\subsection{Performance Evaluation}
We trained the stroke detection algorithm for 100 epochs with early stopping, minimizing binary cross-entropy loss functions. All models were trained with the AdaBound optimizer \cite{adabound}, which used bounds on a dynamic learning rate to transition smoothly from an adaptive method to the more traditional stochastic gradient descent. This approach allowed the model to maintain a higher rate of convergence in early training epochs. The code was written in PyTorch, and experiments were run on an NVIDIA DGX-1.

\section{Results}

\begin{figure*}[h!]
\centerline{\includegraphics[scale=0.7]{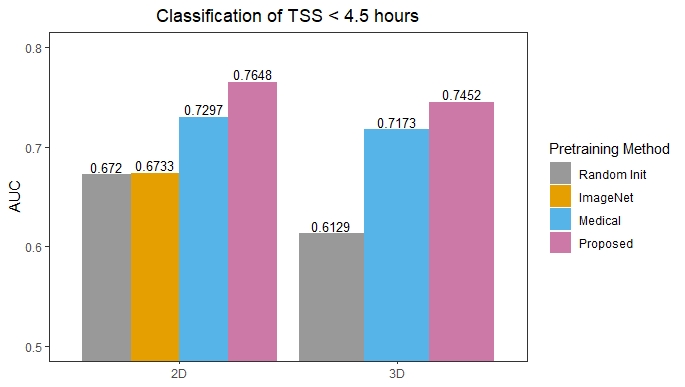}}
\caption{On second phase task TSS $<$ 3 hours, for 2D model, our proposed transfer learning approach has a 5.1\% increase, whereas for the 3D model, there is a 8.3\% increase in ROC-AUC score.} \label{Figure 3}
\end{figure*}

The performance metrics for all of our training phases are summarized in Table \ref{tab:metrics}. For stroke detection, the 2D and 3D architectures achieved ROC-AUC values of 0.8905 and 0.9460, respectively. This indicates that the models were able to reliably identify stroke at both the slice and volume level, which aligns with intensity differences usually observed for stroke lesions on DWI and FLAIR series. For the second training phase, classifying TSS $<$ 3 hours, our pretraining approach improved the performance of 2D model by 14.0\% and our 3D model by 21.6\% when compared to random initialization or to pretraining on natural images (2D model only). For both models, we also examined TSS classification performance with weights pretrained on medical image datasets. We used models trained for brain tumor classification and segmentation to initialize our 2D and 3D models, respectively, given that these tasks are in the same domain and use the same medical imaging modalities \cite{2d_tumor_class_pretrained, 3d_tumor_seg_pretrained}. We froze the weights from earlier layers for both models, and we compared the effect of this pretraining to frozen weights learned from our stroke detection task. While performance improvement was observed using medical image pretraining, our pretraining approach was able to achieve higher performance for both models when compared to both natural image and domain-specific pretraining, with the 2D and 3D models achieving 76.48\% and 74.52\% increase in AUC, respectively. 

\begin{figure*}[h!]
\centerline{\includegraphics[scale=0.7]{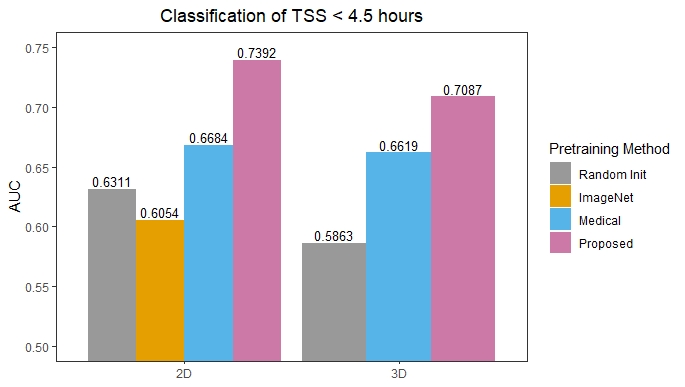}}
\caption{On third phase task TSS $<$ 4.5 hours, for 2D model, our proposed transfer learning approach has a 22.1\% increase in AUC; for 3D model, there is a 20.9\% increase} \label{Figure 3}
\end{figure*}

In the third phase, we train the models to classify TSS $<$ 4.5 hours using weights from the second phase. As shown in the Barchart \label{fig: Figure 3}, both the 2D and 3D models improve by . For the 2D model, pretraining on natural images reduced performance, which has been observed for other medical-image specific tasks \cite{medical_transfer}. As in Phase 2, We also show the results from ImageNet, Tumor detection and segmentation weight transfer for comparison. As expected, due to the similarity of the dataset, the performance improvement is high, from AUC 0.6311 to 0.6684 and from 0.58 to 0.66 for 2D and 3D, respectively. However, the performance improvement (12.9\% and 5.9\%) is still lower than our proposed method (17.1\% and 20.9\%  to AUC 0.7392 and 0.7087). For both tasks, the 2D model achieves higher performance than the 3D model, even with random initialization. 

\begin{table}[htbp]
    \centering
    \begin{tabular*}{1.052\textwidth}{|c|cc|cccc|}
    \hline
    \textbf{Stage} & \textbf{Model} & \textbf{Weights} & \textbf{Sens.} & \textbf{Spec.} & \textbf{Acc.} & {AUC} \\
    \hline
    Phase 1 & 2D & Random & 0.7347 & 0.9286 & 0.8316 & 0.8905 \\
    Stroke Detection & 3D & Random & 0.7732 & 0.9579 & 0.8646 & 0.9460 \\
    \hhline{|=|==|====|}
    Phase 2 & 2D & Random & 0.2444 & 0.9310 & 0.5135 & 0.6720\\
    TSS $<$ 3 hrs & & ImageNet & 0.7879 & 0.5510 & 0.6463 & 0.6733\\
    & & Medical & 0.6970 & 0.7142 & 0.7073 & 0.7297\\
    & & Phase 1 & 0.8222 & 0.6552 & 0.7568 & \textbf{0.7648} \\
    & 3D & Random & 0.7143 & 0.4848 & 0.6220 & 0.6129\\
    & & Medical & 0.5952 & 0.7750 & 0.6829 & 0.7173\\
    & & Phase 1 & 0.8904 & 0.6000& \textbf{0.7724} & 0.7452 \\
    \hhline{|=|==|====|}
    Phase 3 & 2D & Random & 0.2162 & 0.9189 & 0.5676 & 0.6311\\
    TSS $<$ 4.5 hrs & & ImageNet & 0.8789 & 0.4285 & 0.6098 & 0.6054\\
    & & Medical & 0.6666 & 0.6939 & 0.6829 & 0.6684\\
    & & Phase 2 & 0.5405 & 0.7838 & 0.6622 & \textbf{0.7392} \\
    & 3D & Random & 0.3750 & 0.6429 & 0.5122 & 0.5863\\
    & & Medical & 0.8788 & 0.4489 & 0.6220& 0.6619\\
    & & Phase 2 & 0.6279 & 0.7895 & \textbf{0.7037} & 0.7087 \\
    \hline
    Phase 4 & ML & &  0.6522 & 0.7143 & 0.6363 & 0.7174 \\
    TSS $<$ 4.5 hrs  & 2D & Phase 3 & 0.7027 & 0.8108 & \textbf{0.7568} &\textbf{0.7407} \\
    attention+fine-tune & 3D & Phase 3 & 0.5405 & 0.8378 & 0.6892 & 0.7370 \\
    \hline
    DWI-FLAIR  & \multicolumn{2}{l}{Rad 1} & 0.5476 & 0.8500 & 0.6951 & \\
    Mismatch & \multicolumn{2}{l}{Rad 2} & 0.4286 & 0.9250 & 0.6707 & \\
     & \multicolumn{2}{l}{Rad 3} & 0.5714 & 0.6500 & 0.6098 & \\
     & \multicolumn{2}{l}{Agg Rad} & 0.5730 & 0.8750 & 0.7195 & \\

    \hline
    

    \end{tabular*}%
  \caption{\label{tab:metrics} Performance metrics across tasks and architectures. Double lines separate models with different outputs. Sens = Sensitivity, Spec = Specificity, Acc = Accuracy, AUC = Receiver Operating Characteristic Area Under Curve, Rad = Radiologist, Agg Rad = Aggregate Radiologist.}
\end{table}%

In the last of our proposed training phases, fine tuning the attention modules yields improved performance for both the 2D and 3D models, though the improvement was more notable for the 3D model. The optimal ROC-AUC scores for classification of TSS $<$ 4.5 hours are 0.7407 and 0.7370 for 2D and 3D respectively with 17.4\% and 25.7\% performance gain compared to training from scratch.  


\begin{figure}[hbt!]
\centerline{\includegraphics[scale=0.5]{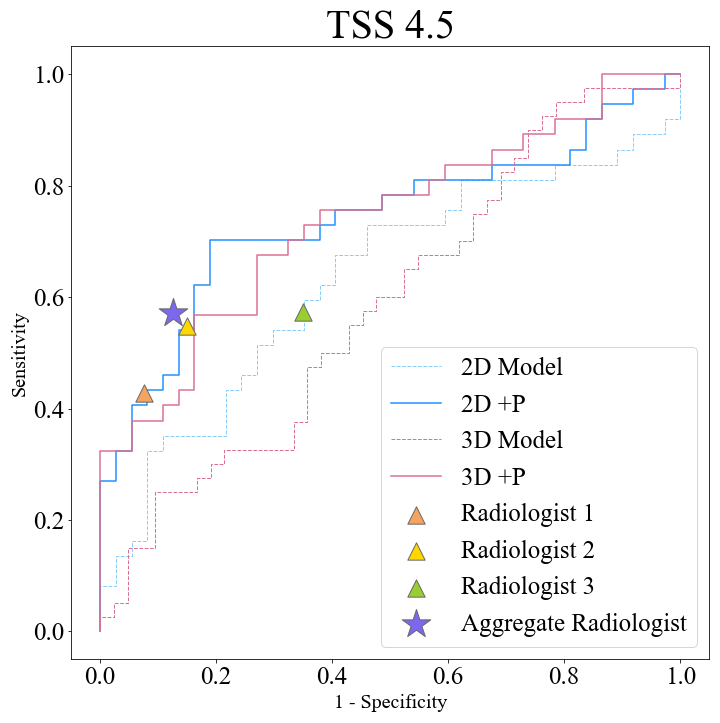}}
\caption{ROC curves for classifying TSS $<$ 4.5 hours. +P = with pretraining.} \label{total_curves}
\end{figure}

For each model, we computed Youden's J statistic and reported the sensitivity, specificity, accuracy, and ROC-AUC score. We compared our model to the performance metrics of each radiologist's DWI-FLAIR mismatch assessments, which served as a proxy for TSS. We also compared our model to the previously-published model with the highest performance metrics by applying this model to our own dataset \cite{tss_korea}; these metrics are included in Table \ref{tab:metrics}. Of note, the inter-reader agreement (Fleiss’ kappa) was 0.46 among all three radiologists, which is typically regarded as a moderate level of agreement and aligns with previous findings of high variability among reader assessments. We also reported the ROC-AUC curves for each of our models in Fig. \ref{total_curves}. 

\begin{figure*}[h!]
\centerline{\includegraphics[scale=0.5]{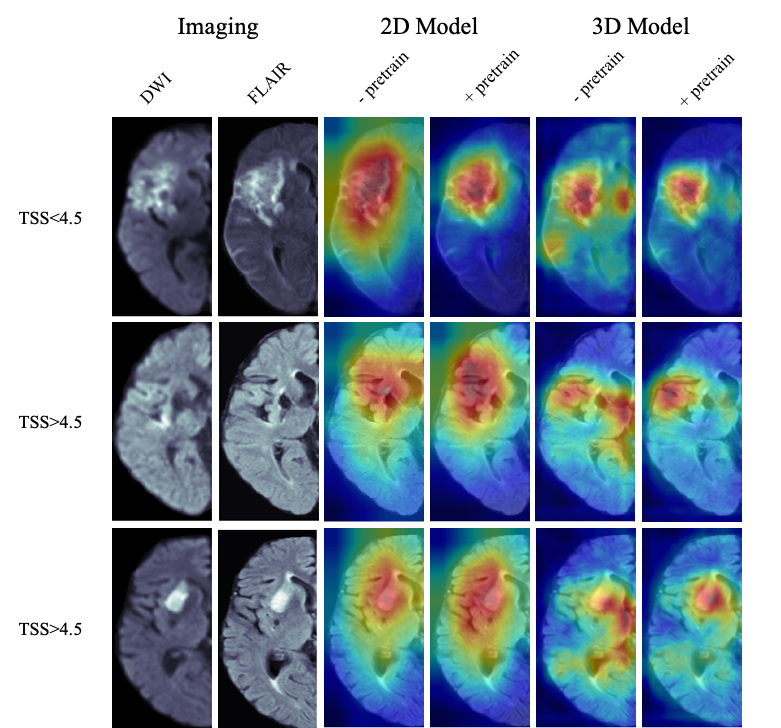}}
\caption{Grad-CAM visualizations of the penultimate convolutional layer for 2D and 3D models, both from scratch and with pretraining.} \label{GradCAM}
\end{figure*}

We generated GradCAMs to visually assess model activation. To evaluate the utility of GradCAMs in a clinical context, an expert radiologist evaluated the overlap of activation map and stroke lesion. The radiologist found that, for slices most representative of stroke lesion, 96\% of cases evaluated had substantial overlap ($>$50\%) between the lesion and activation, while the remainder of cases had moderate overlap. This indicates that Grad-CAM can qualitatively localize to stroke lesions when trained on the TSS tasks.

\section{Discussion}

Among the models tested, the pretrained 2D model achieved the highest performance metrics with a sensitivity of 0.70 and a specificity of 0.81 in classifying TSS $<$ 4.5 hours. Our model was more sensitive than the DWI-FLAIR assessments performed by the neuroradiologists, which we treated as a surrogate for determining a TSS $<$ 4.5 hours. We also compared our model to the previously published state-of-the-art method. The threshold method implemented in \cite{tss_korea}, which was used to create the ROI, was very stringent, in that only 221 of our original 422 patients had large enough ROI from which features could be extracted. Thus, their performance metrics represent a subset of our larger dataset. We also tested our model performance on this subset and achieved an ROC-AUC of 0.76. Nevertheless, on the entire dataset, the optimal 2D model with pretraining was able to outperform the previous model. From a clinical perspective, these results indicate that our model may be able to correctly identify more patients within the 4.5 hour window and therefore eligible to receive thrombolytic therapy when compared to both DWI-FLAIR mismatch assessment and the threshold-based machine learning method. There are many tasks within the medical image domain to which our proposed task-adaptive pretraining schema can be applied. For example, this schema could be used for brain tumor classification, where brain tumor detection is the pretraining task. 

The optimal 2D model has a ROC-AUC comparable to that of the 3D model; however, the sensitivity (0.54) and specificity (0.84) of the 3D model are less balanced, indicating that while the rate of true negatives is high, there are less true positives identified by that model. In total, our model metrics illustrate that the progressive pretraining schema enhances performance for our task considerably, regardless of the model architecture. For both models, attention modules enhance the performance. During training, the models with attention modules are less likely to over-fit. The use of GradCAM for our models highlights regions of the brain that impact decisions, as illustrated in Fig.\ref{GradCAM}. The GradCAMs illustrate that the pretrained model is able to more precisely localize to the stroke infarct and highlight other regions outside of the infarct that may inform this classification task. 

Our model performance metrics are comparable to previous approaches in TSS classification. However, this study has a few factors that increase its potential clinical applicability. The patients in our dataset comprise a wider range of stroke locations and other clinical demographics than in previously assessed datasets. Additionally, our model leverages the entire brain hemisphere, which may contain more relevant information among this broader patient cohort. This has the potential to reduce bias in our model, and with the convolutional architecture, allows this information to be incorporated in decision-making.

That said, deep learning generally requires a high volume of data. While many medical image-related tasks have used deep learning with a comparable amount of patient data used here, a higher volume of data would greatly enhance the model performance. This model only uses diffusion-based imaging, as these are the images used in current clinical practice. Incorporating perfusion-based imaging and its derivatives such as perfusion maps may better inform TSS. There is a substantial body of work using perfusion imaging parameters for stroke outcomes \cite{perf_1,perf_2,tss_kch_1,tss_key}. Finally, the use of clock time as a label for TSS may not fully encompass the physiology underlying ischemia in the brain; for example, cerebral collateral flow may compensate for a hypoperfused area within the brain and reduce the amount of ischemia that tissue is experiencing during a stroke \cite{collaterals}, which may be the biological reason for DWI-FLAIR mismatch.

\section{Conclusion}

This approach uses 2D and 3D CNN models to classify TSS for 422 patients and compares model performances to DWI-FLAIR mismatch readings performed by three expert neuroradiologists. We demonstrate that our 2D model outperforms the 3D model when classifying TSS $<$ 4.5 hours, which is the current clinical guideline. We show that pretraining the model on stroke detection, then refining the model on TSS classification yields better performance than training on TSS classification labels alone; the incorporation of soft attention modules also enhances performance of both the 2D and 3D  when compared to CNNs without them. By visualizing network gradients via Grad-CAM, we show that our pretrained models localize to stroke infarcts and surrounding regions. We demonstrate that our both our 2D and 3D model is able to generalize to an inclusive dataset comprising multiple types of ischemic stroke, and that this model may be able to inform TSS for patients with unknown symptom onset. 

\section{Acknowledgments}
This work was supported by the following grants: NIH T32EB016640-07, NIH R01NS100806-02, and NVIDIA Academic Hardware Grant. The content is solely the responsibility of the authors and does not necessarily represent the official views of the National Institutes of Health.



%
%
\bibliographystyle{elsarticle-num}
\bibliography{mybibfile.bib}
\end{document}